\documentclass[A4,11pt,epsf,oneside]{article}
\usepackage{amssymb, amscd, amsmath,amsthm,amsfonts}
\usepackage{graphicx}
\usepackage{epsfig}
\usepackage{color}
\usepackage{capt-of}
\usepackage{float}
\usepackage{cite}

\begin{document}
%\doublespace
%\setlength{\baselineskip}{17pt} (for double line space)
\renewcommand{\thefootnote} {\fnsymbol{footnote}}
\setcounter{page}{1}

\title{\textbf{Dynamics of Particles around Time Conformal Schwarzschild Black Hole}}
\author{\bf{Abdul
Jawad}$^1$\thanks{jawadab181@yahoo.com;
abduljawad@ciitlahore.edu.pk}, \bf{Farhad
Ali}$^{2}$\thanks{farhadmardan@gmail.com}, \bf{M.Umair Shahzad}$^1$
\thanks{m.u.shahzad@ucp.edu.pk}\\ and \bf{G. Abbas}$^3$ \thanks{abbasg91@yahoo.com}\\
$^1$Department of Mathematics, COMSATS Institute of Information\\
Technology, Lahore-54000, Pakistan.\\
$^2$Department of Mathematics, Kohat University of Science and\\
Technology, Kohat, Pakistan.\\
$^3$Department of Mathematics, The Islamia University of\\
Bahawalpur, Bahawalpur Pakistan.}

 \maketitle
\begin{abstract}
In this work, we present the new technique for discussing the
dynamical motion of neutral as well as charged particles in the
absence/presence of magnetic field around the time conformal
Schwarzschild black hole. Initially,  we find the numerical
solutions of geodesics of Schwarzschild black hole and the time
conformal Schwarzschild black hole. We observe that the
Schwarzschild spacetime admits the time conformal factor
$e^{\epsilon f(t)}$, where $f(t)$ is an arbitrary function and
$\epsilon$ is very small which causes the perturbation in the
spacetimes. This technique also re-scale the energy content of
spacetime. We also investigate the thermal stability, horizons and
energy conditions corresponding time conformal Schwarzschild
spacetime. Also, we examine the dynamics of neutral and charged
particle around time conformal Schwarzschild black hole. We
investigate the circumstances under which the particle can escape
from vicinity of black hole after collision with another particle.
We analyze the effective potential and effective force of particle
in the presence of magnetic field with angular momentum graphically.
\end{abstract}

\section{Introduction}

The dynamics of particles (massless or massive, neutral or charged)
in the vicinity of black hole (BH) is the most intriguing problems
in the BH astrophysics. The geometrical structure of spacetime in
the surrounding of BH could be studied better by these phenomenon
\cite{1,2}. The motion of particles help us to understand
gravitational fields of BHs experimentally and to compare them with
observational data. The magnetic field in the vicinity of BH is due
to the presence of plasma \cite{3}. Magnetic fields can't change the
geometry of BH but its interaction with plasma and charged particles
is very important \cite{4,5}. The transfer of energy to particles
moving around BH geometry is due to magnetic field, so that there is
a possibility of their escape to spatial infinity \cite{6}. Hence,
the high energy may produce by charged particles collision in the
presence of magnetic field rather than its absence.

Recent observations have also provided hints about connecting
magnetars with very massive progenitor stars, for example an
infrared elliptical ring or shell was discovered surrounding the
magnetar SGR $1900+14$ \cite{6a}. The explanation of magnetic white
dwarfs was first proposed in the scenario of fossil-field for
magnetism of compact objects \cite{6b,6c,6d}. Magnetic white dwarfs
may be created as a result of rebound shock explosion \cite{6d} and
may further give rise to novel magnetic modes of global stellar
oscillations \cite{6e}. This fossil-field scenario is supported by
the statistics for the mass and magnetic field distributions of
magnetic white dwarfs. Magnetized massive progenitor stars with a
quasi-spherical general polytropic magneto fluid under the
self-gravity are modeled by \cite{6f}. Over recent years methods are
being developed to detect truly cosmologically magnetic fields, not
associated with any virilized structure. The spectral energy
distribution of some tera electron-volt range (TeV) blazers in the
TeV and giga electron volt (GeV) range hint at the presence of a
cosmological magnetic field pervading all space \cite{6g,6h}.

Authors \cite{11,12,13} investigated effects on charged particles in
the presence of magnetic field near BHs. The motion of charged
particles discussed by Jamil et al. \cite{14} around weakly
magnetized BHs in JNW space time. Many aspects of motion of
particles around Schwarzschild and Reissner-Nordstrom BH have been
studied and the detail review is given the reference \cite{15,16}.
An important question may rise in the studies of these important
problems is "what will happen in the dynamics of particles (neutral
or charged) around BHs with respect to time". Particles moves in
circular orbits around BHs in equatorial plane may show drastic
change with respect to time. Following the work done by Zaharani et
al. \cite{16}, we present at what conditions particle escape to
infinity after collision. At some specific time interval, we study
the dynamics of neutral and charged particles around time conformal
Schwarzschild BH.

The purpose of our work is to investigate under which circumstances
the particle moving initially would escape from innermost stable
circular orbit (ISCO) or it remain bounded, captured by BH or escape
to infinity after collision with another particle. We calculate the
escape velocity of particle and investigate some important
characteristics such as effective potential and effective force of
particle motions around the BH with respect to time. Initially, we
will find attraction in particle motion but after some specific time
we see the repulsion. The comparison of stability orbits of particle
with the help of Lyapunov exponent is also established.

The outline of paper is as follows: In section \textbf{2}, we
discuss the Noether and Approximate Noether Symmetries for
Schwarzschild BH and time conformal Schwarzschild BH and find the
corresponding conservation laws. In section \textbf{3}, we
investigate the dynamics of neutral particle through effective
potential, effective force and escape velocity. In section
\textbf{4}, the motion of charged particle is discussed, the
behavior of effective potential, effective force and escape velocity
is analyzed in the presence of magnetic field. In section
\textbf{5}, the Lyapunov exponent is discussed. Conclusion and
observations are given in the last section.

\section{Noether and Approximate Noether Symmetries and Corresponding Conservation Laws}

The one to one correspondence between conservation laws and
symmetries of Lagrangian (Noether symmetries) has been firstly
pointed out by Emmy Noether \cite{8,9,10}. She suggested that there
exist a conservation law for every Noether symmetry. One can obtain
the approximate Noether symmetries of Schwarzschild solution by
considering its first order perturbation and investigate the
energy-momentum of corresponding spacetime.

\subsection{Schwarzschild Black Hole}

The line element of the Schwarzschild solution is\begin{equation}
ds^{2}_e=(1-\frac{2M}{r})dt^{2}-(1-\frac{2M}{r})^{-1}dr^{2}-
r^2({d\theta}^{2}+\sin\theta{d\phi}^{2}).\label{1}\end{equation}
the corresponding Lagrangian
is\begin{equation}L_e=(1-\frac{2M}{r})\dot{t}^{2}-(1-\frac{2M}{r})^{-1}\dot{r}^{2}
-r^2(\dot{\theta}^{2}+\sin\theta\dot{\phi}^{2}).\label{2}\end{equation}
The symmetry generator
\begin{equation}\textbf{X}^{1}_e=\xi_e\frac{\partial}{\partial s}
+\eta^{i}_e\frac{\partial}{\partial
x^i}+\eta^i_{es}\frac{\partial}{\partial \dot{x}^i},\quad i=0,1,2,3
\label{6}\end{equation} is the first order prolongation of
\begin{equation}\textbf{X}_e=\xi_e\frac{\partial}{\partial s}
+\eta^{i}_e\frac{\partial}{\partial x^i}.\label{7}\end{equation}
$\textbf{X}_e$ is the Noether symmetry if it satisfies the equation,
\begin{equation} \textbf{X}^{1}_eL_e+(D\xi_e)L_e=DA_e,\label{8}
\end{equation} where $D$ is differential operator of the form
\begin{equation}D=\frac{\partial}{\partial
s}+\dot{x}^i\frac{\partial}{\partial x^i},\label{9}\end{equation}and
$A_e$ is Gauge function.

The solution of the system (\ref{8}) is
\begin{equation}
\begin{split}
& A_e ={C_2},\quad \eta_e^0
=C_3,\\&\eta^1_e  =0,\quad\eta^2_e  = -C_5\cos\phi +C_6\sin\phi
,\\&\eta^3_e =C_4+ \frac {\cos \theta  \left( C_5\sin\phi +C_6\cos
\phi  \right) }{\sin \theta },\\&\xi^0 =C_1.
\label{exc}
\end{split}
\end{equation}

The corresponding Noether symmetries generators are
\begin{equation}\begin{split}& \mathbf{X_1}=\frac{\partial}{\partial
t}, \quad \mathbf{X_2}=\frac{\partial}{\partial s}, \quad
\mathbf{X_3}=\frac{\partial}{\partial \phi}, \\&
\mathbf{X_4}=\cos\phi\frac{\partial}{\partial\theta}-\cot\theta\sin\phi\frac{\partial}{\partial\phi},
\quad
\mathbf{X_5}=\sin\phi\frac{\partial}{\partial\theta}+\cot\theta\cos\phi\frac{\partial}{\partial\phi}.
\label{0138}\end{split}\end{equation} For conservation laws (given
in the following table), we use the following equation
\begin{align}\phi=\frac{\partial\mathcal{L}}{\partial\dot{x}^i}(\eta^i-\xi\dot{x}^i)+\xi\mathcal{L}-A.\label{con}\end{align}
For example, let us see the calculation for symmetry $\mathbf{X_1}$.
In this symmetry, $\xi=0$ and $\eta^0=1$. Also from lagrangian, we
have
$\frac{\partial\mathcal{L}}{\partial\dot{t}}=2(1-\frac{2M}{r})\dot{t}$.
Putting these values in equation (\ref{con}) we have
$$\phi=2(1-\frac{2M}{r})\dot{t}(1-0)+0-0=2(1-\frac{2M}{r})\dot{t}.$$
Similarly for other symmetries, the conservation laws are given in
the table
\begin{table}[H]
\begin{center}\captionof{table}{First Integrals}
\begin{tabular}{|p{1.7cm}|p{10cm}|}
 \hline
 Generators &First Integrals\\
                \hline
               $ \mathbf{X_1}$ & $\phi_1=2\big(1-\frac{2M}{r}\big)\dot{t} $ \\
               $\mathbf{X_2}$&$\phi_2=-\bigg(\big(1-\frac{2M}{r}\big)\dot{t}^2-\big(1-\frac{2M}{r}\big)^{-1}\dot{r}^2-r^2 \left(\dot{\theta}^2+\sin^2
               \theta\dot{\phi}^2 \right)\bigg)$\\

               $ \mathbf{X_3} $& $\phi_3=-2r^2\sin^2\theta\dot{\phi}$ \\

               $ \mathbf{X_4}$ & $\phi_4=-2r^2\left (\cos\phi\dot{\theta} -\cot\theta\sin\phi\dot{\phi} \right)$ \\

                $\mathbf{X_5}$ & $\phi_5=-2r^2\left (\sin{\phi}\dot{\theta}+\cot\theta\cos\phi \dot{\phi} \right)$\\
               \hline
\end{tabular}\end{center}\end{table}

\section{Perturbed Metric}

Perturbing the metric given by (\ref{1}) by using the general time
conformal factor $e^{\epsilon f(t)}$, which
gives\begin{equation}ds^{2}=e^{\epsilon f(t)}ds^{2}_e=(1+\epsilon
f(t)+\frac{\epsilon^2
f(t)^2}{2}+.....)ds^2_e,\label{3}\end{equation} Picking the first
order terms in $\epsilon$ and neglecting the higher order terms we
have\begin{equation}ds^{2}=e^{\epsilon f(t)}ds^{2}_e=(1+\epsilon
f(t))ds^2_e=ds^2_e+\epsilon f(t)ds^2_e,\label{3}\end{equation}in
expanded form \begin{equation}\begin{split}
&ds^{2}=(1-\frac{2M}{r})dt^{2}-(1-\frac{2M}{r})^{-1}dr^{2}-r^2({d\theta}^{2}+\sin\theta{d\phi}^{2})+\\&\epsilon
f(t)\bigg((1-\frac{2M}{r})dt^{2}-(1-\frac{2M}{r})^{-1}dr^{2}-r^2({d\theta}^{2}+\sin\theta{d\phi}^{2})\bigg),\\&
L=(1-\frac{2M}{r})\dot{t}^{2}-(1-\frac{2M}{r})^{-1}\dot{r}^{2}-r^2(\dot{\theta}^{2}+\sin\theta\dot{\phi}^{2})+\\&\epsilon
f(t)\bigg((1-\frac{2M}{r})\dot{t}^{2}-(1-\frac{2M}{r})^{-1}\dot{r}^{2}-r^2(\dot{\theta}^{2}+\sin\theta\dot{\phi}^{2})\bigg),
\label{5}\end{split}\end{equation}where $`` \ \dot{} \ "$ denotes
the differentiation with respect to $s$, $L_e$ is defined in
equation (\ref{2}) and
\begin{align*}L_a=f(t)\bigg((1-\frac{2M}{r})\dot{t}^{2}-(1-\frac{2M}{r})^{-1}\dot{r}^{2}-r^2(\dot{\theta}^{2}+\sin\theta\dot{\phi}^{2})\bigg).\end{align*}
We define the first order approximate Noether symmetries \cite{7}
\begin{equation}\mathbf{X}=\mathbf{X}_e+ \epsilon
\mathbf{X}_a,\label{10}\end{equation} up to the gauge
   $ A=A_e+\epsilon A_a.$ Where\begin{equation}\mathbf{X}_a=\xi_a\frac{\partial}{\partial s}+\eta^{i}_a\frac{\partial}{\partial x^i}, \quad i=4,5,6,7\label{11}\end{equation}is the approximate Noether symmetry and $A_a$ is the approximate part of the gauge function.

Now $\textbf{X}$ is the first order approximate Noether symmetry if
it satisfy the equation
\begin{equation}
\mathbf{X}^1L+(D\xi)L=DA,\label{12}
\end{equation} where $\mathbf{X}^1$
is the first order prolongation of the first order approximate
Noether symmetry $\textbf{X}$ given in equation (\ref{10}).

The equation (\ref{12}) split into two parts that is
\begin{equation}\mathbf{X}_eL_e+(D\xi_e)L_e=DA_e,\label{13}\end{equation}
\begin{equation}\textbf{X}^{1}_aL_e+\textbf{X}^{1}_eL_a+(D\xi_e)L_a+(D\xi_a)L_e=DA_a.\label{14}\end{equation}

All $\eta^{i}_e,$ $\eta^{i}_a,$ $\xi_e,$ $\xi_a,$ $ A_e$ and $A_a$
are the functions of $s,t,r,\theta,\phi$ and $\dot{\eta^{i}_e},$
$\dot{\eta^{i}_a}$ are functions of
$s,t,r,\theta,\phi,\dot{t},\dot{r},\dot{\theta},\dot{\phi}$. From
equation (\ref{14}) we obtained a system of 19 partial differential
equations whose solution will provide us the cases where the
approximate Noether symmetry(ies) exist(s). By putting the exact
solution given in equation (\ref{exc}) in  equation (\ref{14}) we
have the following system of 19 PDEs
\begin{equation}\begin{split}&\xi^1_t=\xi^1_r=\xi^1_\theta=\xi^1_\phi=A_{as}=0,\quad
2\eta^4_s(1-\frac{2M}{r})-A_{at}=0,\quad
2\eta^5_s(1-\frac{2M}{r})^{-1}+A_{ar}=0,\\&2\eta^6_sr^2+A_{a\theta}=0,\quad
2\eta^7_sr^2\sin^2\theta+A_{a\phi}=0,\quad
\eta^4_\phi(1-\frac{2M}{r})-r^2\sin^2\theta\eta^7_t=0,\\&\eta^4_\theta(1-\frac{2M}{r})-r^2\eta^6_t=0,\quad
\eta^4_r(1-\frac{2M}{r})^2-\eta^5_t=0,\quad
\eta^5_\phi(1-\frac{2M}{r})^{-1}-r^2\sin^2\theta\eta^7_r=0,\\&\eta^5_\theta(1-\frac{2M}{r})^{-1}-r^2\eta^6_r=0,\quad
\eta^6_\phi-\sin^2\theta\eta^7_\theta=0,\quad
C_3f_t(t)+\frac{2}{r}\eta^5+2\eta^5_r-\xi^1_s=0,\\&C_3f_t(t)-\frac{2M}{r^2(1-\frac{2M}{r})}\eta^5+2\eta^5_r-\xi^1_s=0,\quad
 C_3f_t(t)+\frac{2M}{r^2(1-\frac{2M}{r})}\eta^5+2\eta^5_r-\xi^1_s=0,\\&C_3f_t(t)+\frac{2}{r}\eta^5+2\cot\theta\eta^6+2\eta^7_\phi-\xi^1_s=0.
\label{apr}\end{split}\end{equation}We see that the system
(\ref{apr}) have $C_3$ and $f(t)$ in it. The solution of this system
(\ref{apr}) is\begin{equation}\begin{split}& A_a ={C_2}^{'},\quad
\eta_a^4 =C_3^{'},\\&\eta^5_a  =0,\quad\eta^6_a  = -C_5^{'}\cos\phi
+C_6^{'}\sin\phi ,\\&\eta^7_a =C_4^{'}+ \frac {\cos \theta  \left(
C_5^{'}\sin\phi
  +C_6^{'}\cos \phi  \right) }{\sin
\theta },\\&\xi^0 =C_1^{'}+C_3\frac{s}{\alpha}, \quad
f(t)=\frac{t}{\alpha}. \label{far}\end{split}\end{equation}
combining the solution (\ref{exc}) and (\ref{far}) we have the
following solution of equation
(\ref{12})\begin{equation}\begin{split}& A_e+\epsilon A_a
={C_2}+\epsilon C_2^{'},\quad \eta_e^0+\epsilon\eta^4_a
=C_3+\epsilon C_3^{'},\\&\eta^1_e+\epsilon\eta^{{5}}_a
=0,\quad\eta^2_e+\epsilon\eta^{{6}}_a  = -(C_5+\epsilon C_5^{'})\cos
\left( \phi \right) +(C_6+\epsilon C_6^{'})\sin \left( \phi
 \right) ,\\&\eta^3_e+\epsilon\eta^{{7}}_a  =(C_4+\epsilon C_4^{'})+
\frac {\cos \left( \theta \right)  \left( (C_5+\epsilon C_5^{'})\sin
\left( \phi
 \right) +(C_6+\epsilon C_6^{'})\cos \left( \phi \right)  \right) }{\sin \left(
\theta \right) },\\&\xi^0+\epsilon\xi_{{1}} =\epsilon{\frac {
 C_3s}{\alpha}}+(C_1+\epsilon C_1^{'}),\quad f(t)=\frac{t}{\alpha}.
\end{split}\end{equation}In symmetry generators form we have
\begin{equation}\begin{split}&
\mathbf{X_1}=\frac{\partial}{\partial
t}+\epsilon\frac{s}{\alpha}\frac{\partial}{\partial s}, \quad
\mathbf{X_2}=\frac{\partial}{\partial s}, \quad
\mathbf{X_3}=\frac{\partial}{\partial \phi}, \\&
\mathbf{X_4}=\cos\phi\frac{\partial}{\partial\theta}-\cot\theta\sin\phi\frac{\partial}{\partial\phi},
\quad
\mathbf{X_5}=\sin\phi\frac{\partial}{\partial\theta}+\cot\theta\cos\phi\frac{\partial}{\partial\phi}.
\label{0138}\end{split}\end{equation}We see that only the symmetry
$\mathbf{X_1}$ got the non-trivial approximate part which re-scale
the energy content of the Schwarzschild spacetime. The corresponding
conservation laws are \begin{table}[H]
\begin{center}\captionof{table}{First Integrals}
\begin{tabular}{|p{.5cm}|p{11cm}|}
\hline
Gen&First Integrals\\
\hline
               $ \mathbf{X_1}$ & $E_{approx}=2\big(1-\frac{2M}{r}\big)\dot{t}+\frac{\epsilon}{\alpha}\bigg(2t\dot{t}\big(1-\frac{2M}{r}\big)-sL\bigg) $ \\
               $\mathbf{X_2}$&$Lag=(1+\frac{\epsilon t}{\alpha})\bigg(\big(1-\frac{2M}{r}\big)\dot{t}^2-\big(1-\frac{2M}{r}\big)^{-1}\dot{r}^2-r^2 \left(\dot{\theta}^2+\sin^2
               \theta\dot{\phi}^2 \right)\bigg)$\\

               $ \mathbf{X_3} $& $-L_z=(1+\frac{\epsilon t}{\alpha})r^2\sin^2\theta\dot{\phi}$ \\

               $ \mathbf{X_4}$ & $\phi_4=-(1+\frac{\epsilon t}{\alpha})r^2\left (\cos\phi\dot{\theta} -\cot\theta\sin\phi\dot{\phi} \right)$ \\

                $\mathbf{X_5}$ & $\phi_5=-(1+\frac{\epsilon t}{\alpha})r^2\left (\sin{\phi}\dot{\theta}+\cot\theta\cos\phi \dot{\phi} \right)$\\
               \hline
\end{tabular}\end{center}\end{table}

\subsection{Existence and Location of Horizon}

Consider the line element of time conformal Schwarzschild BH
\begin{equation}\label{c1}
ds^{2}=\left(1+\frac{\epsilon
t}{\alpha}\right)\left(\left(1-\frac{2M}{r}\right)dt^{2}-\left(1-\frac{2M}{r}\right)^{-1}dr^{2}-r^2({d\theta}^{2}
+\sin\theta{d\phi}^{2})\right).
\end{equation}
The parameter $\epsilon$ is a dimensionless small parameter that
causes the perturbation in the spacetime and $\alpha$ is constant of
the dimension equal to that of time $t$ so that to make the term
$\frac{t}{\alpha}$ as dimensionless. $M$ is the mass of the black
hole. In order to discuss the trapped surfaces and apparent horizon
of the above metric, we use the definitions of \cite{R1}-\cite{R4}
and Eqs.(13) and (14) of \cite{R4}. Here, we define the mean
curvature one-form as
\begin{equation}\label{R1}
H_{\mu}=\delta^{a}_{\mu}(U,{\mu}-div{\vec{g}}_{a}).
\end{equation}
Also, the scalar defining the trapped surface of given spacetime is
\begin{equation}\label{R2}
\kappa=-g^{bc}H_{b}H_{c}.
\end{equation}
In the notation of \cite{R4}, the quantities in the above equations
are defined as $G\equiv e^{U}=\sqrt{detg_{AB}}$ and
${\vec{g}}_{a}=g_{aA}dx^{A}$. The coordinates are defined by
$\{x^a\}=\{t,r\}$ and $\{x^A\}=\{\theta,\phi\}$. By using these
coordinates, we obtained
\begin{eqnarray}\label{R3}
&&{\vec{g}}_{t}=0 , \quad\quad\quad{\vec{g}}_{r}=0\\\label{R4}
&&e^{U}=\sqrt{detg_{AB}}=(1+\frac{\epsilon t}{\alpha})r^2 sin\theta.
\end{eqnarray}
Hence, using given metric (\ref{c1}), Eqs.(\ref{R1})-(\ref{R4}), we
get following form of trapping scalar
\begin{equation}\label{R5}
\kappa=\frac{-(\frac{\epsilon}{\alpha})^2r^2+4(1-\frac{2m}{r})^2(1+\frac{\epsilon
t}{\alpha})^2}{(1-\frac{2m}{r})(1+\frac{\epsilon t}{\alpha})^3r^2}.
\end{equation}

Now the surface of the given geometry (\ref{c1}), will be trapped,
marginally trapped and absolutely non-trapped if $\kappa$ is
positive, zero and negative respectively. In order to analyze the
nature of trapped surfaces and location of horizons, we solve
Eq.(\ref{R5}) for $r$ by imposing the restriction on $\kappa$ such
that  $\kappa>0$, $\kappa=0$ and $\kappa<0$. In the following we
discuss these situations in detail.
\begin{itemize}
\item  $\kappa>0$, leads to two positive real values of $r$, $r_+>\frac{(\alpha+\epsilon
t)\Big(1+\sqrt{1-\frac{4m\epsilon}{(\alpha+\epsilon
t)}}\Big)}{\epsilon}$ and $r_-<\frac{(\alpha+\epsilon
t)\Big(1-\sqrt{1-\frac{4m\epsilon}{(\alpha+\epsilon
t)}}\Big)}{\epsilon}$ such that $\alpha>\epsilon(4m-t)$. Here $r_+$
and $r_-$ corresponds to outer and inner horizons for trapping
surface respectively.
\item  $\kappa=0$, leads to two positive real values of $r$, $r_+=\frac{(\alpha+\epsilon
t)\Big(1+\sqrt{1-\frac{4m\epsilon}{(\alpha+\epsilon
t)}}\Big)}{\epsilon}$ and $r_-=\frac{(\alpha+\epsilon
t)\Big(1-\sqrt{1-\frac{4m\epsilon}{(\alpha+\epsilon
t)}}\Big)}{\epsilon}$ such that $\alpha>\epsilon(4m-t)$. Here $r_+$
and $r_-$ corresponds to outer and inner horizons for marginally
trapping surface respectively.
\item $\kappa<0$, leads to two positive real values of $r$, $r_+<\frac{(\alpha+\epsilon
t)\Big(1+\sqrt{1-\frac{4m\epsilon}{(\alpha+\epsilon
t)}}\Big)}{\epsilon}$ and $r_->\frac{(\alpha+\epsilon
t)\Big(1-\sqrt{1-\frac{4m\epsilon}{(\alpha+\epsilon
t)}}\Big)}{\epsilon}$ such that $\alpha>\epsilon(4m-t)$. Here $r_+$
and $r_-$ corresponds to outer and inner absolutely non-trapping
points on the given surface respectively. We would like to mention
that we have considered the denominator of Eq.(\ref{R5}) as positive
for the arbitrary value of parameters and coordinates.
\end{itemize}

\subsection{Thermal Stability}

The time conformal Schwarzschild metric is
\begin{equation}
ds^{2}=\left(1+\frac{\epsilon
t}{\alpha}\right)\left(f(r)dt^{2}-(f(r))^{-1}dr^{2}-r^2({d\theta}^{2}
+\sin\theta{d\phi}^{2})\right),
\end{equation}
where $f(r)=1-\frac{r_{\ast}}{r}$ with $r_{\ast}=2M$. Further
suppose that $v(r,t)=\left(1+\frac{\epsilon t}{\alpha}\right)f(r)$.
Clearly, for values of $r>r_{\ast}$ this solution is positive
definite and coordinate singularity occurs at $r=r_{\ast}$. The
coordinate $t$ is identified periodically with period
\begin{equation}
\beta=\frac{4 \pi}{v'(r,t)}|_{r=r_{\ast}}=\left(1-\frac{\epsilon
t}{\alpha}\right)(4 \pi r_{\ast}).
\end{equation}
In the limit of large $r$, the killing vector
$\frac{\partial}{\partial t}$ is normalized to $1$. The temperature
measure to infinity may be formally identified with inverse of this
period. Hence by Tolman law, for any self gravitating system in
thermal equilibrium, a local observer will measure a local
temperature T which scales as $g_{11}^{-\frac{1}{2}}$ \cite{17a}.
The constant of proportionality in present context is
\begin{equation}
T_{\infty}=\frac{1}{\beta}=\left(1+\frac{\epsilon
t}{\alpha}\right)(4 \pi r_{\ast})^{-1}.
\end{equation}
The wall temperature $T_W$ and surface area $A_W=4\pi r_W^2$ is
defined by York \cite{17}. One topologically regular solution to
Einstein equation with these boundary conditions is hot flat space
with uniform temperature $T_W$. Also another solution is
Schwarzschild metric. If a BH of horizon $r_{\ast}<r_W$ does exist
then the wall temperature from Tolman Law must satisfy
\begin{equation}\label{t1}
T_W=\left(1+\frac{\epsilon t}{\alpha}\right)(4 \pi
r_{\ast})^{-1}\left(1-\frac{r_{\ast}}{r_W}\right)^{-\frac{1}{2}}.
\end{equation}
In terms of $r_W$ and $T_W$, this equation may be solved for
$r_{\ast}$. There is no real positive root for $r_{\ast}$, if $r_W
T_W<\frac{\sqrt{27}}{8 \pi}$ \cite{17a}.

For any value of $r_W$ and $T_W$, the entropy of BH solution to Eq.
(\ref{t1}) is $S=\pi r_{\ast}^2$. The heat capacity of constant
surface for any solution is
\begin{equation}
C_A=T_W \frac{\partial S}{\partial T_W}|_{A_W}=-2 \pi r_{ast}^2
\left(1-\frac{r_{\ast}}{r_W}\right)\left(1-\frac{3r_{\ast}}{2r_W}\right)^{-1}.
\end{equation}
The heat capacity is positive and equilibrium configuration is
locally thermally stable if $r_{\ast}<r_W<\frac{3r_{\ast}}{2}$.

\subsection{Energy Conditions}

Consider the line element of time conformal Schwarzschild BH of Eq.
(\ref{c1}). Let's suppose that the matter distribution is isotropic
in nature, whose energy-momentum tensor is given by
\begin{equation}
T_{\mu \nu}= (\rho + p)u_{\mu} u_{\nu}-p g_{\mu \nu}
\end{equation}
Here the vector $u_i$ is fluid four velocity with
$u_i=(\sqrt{g_{11}},0,0,0)$, $\rho$ is the matter density and
\emph{p} is the pressure.

Taking $G=1=c$, from Einstein field equations one can deduce that
\begin{eqnarray}\label{e1}
8 \pi \rho &=& \frac{\epsilon t}{\alpha r^2} \\
   8 \pi p&=& -\frac{\epsilon t}{\alpha r^2}\label{e2}
\end{eqnarray}
Now, we want to check the energy conditions for time conformal
Schwarzschild BH. For Null Energy Condition (NEC), Weak Energy
Condition (WEC), Strong Energy Condition (SEC) and Dominant Energy
Condition (DEC), the following inequalities must satisfy

NEC: $\rho + p \geq 0$.

WEC: $\rho + p \geq 0$, $\rho \geq 0$.

SEC: $\rho + p \geq 0$, $\rho + 3p \geq 0$.

DEC:  $\rho \geq |p|$.

From Eqs. (\ref{e1}) and (\ref{e2}), we can see that the NEC, WEC
and DEC satisfy by time conformal Schwarzschild BH, on the other
hand, SEC violates by it.

\section{Dynamics of Neutral Particle}

We discuss the dynamics of neutral particle around time conformal
Schwarzschild BH defined by (\ref{1}). The approximate energy
$E_{approx}$ and the approximate angular momentum $L_{z}$ are given
in Table \textbf{2} . The total angular momentum in $(\theta, \phi)$
plane can be calculated from $\phi_4$ and $\phi_5$ which has the
value
\begin{align*}&E_{approx}=(1+\frac{\epsilon t}{\alpha})(E-\frac{\epsilon
s}{\alpha}),\\
&L^2=(1+\frac{\epsilon
t}{\alpha})\bigg(r^2v_{\bot}+\frac{L_z^2}{\sin^2\theta}\bigg),\end{align*}
where
\begin{align}v_{\bot}=r^2\dot{\theta}^2,\quad E=\big(1-\frac{2M}{r}\big)\dot{t},\quad \dot{x}^\mu\dot{x}_\mu=1.\label{n}\end{align}
Using the normalization condition given in Eq. (\ref{n}), we can get
the approximate equation of motion of neutral particle
\begin{align}\dot{r}^2=E^2-\big(1-\frac{2M}{r}\big)\bigg((1-\frac{\epsilon t}{\alpha})+\frac{L_z^2(1-\frac{2\epsilon t}{\alpha})}{r^2\sin^2\theta}\bigg).\label{m}\end{align}
For $\dot{r}=0$ and $\theta=\frac{\pi}{2}$, effective potential
turns out to be

\begin{align}
{E}^2=\big(1-\frac{2M}{r}\big)\bigg(1-\frac{\epsilon t}{\alpha}
+\frac{L_z^2(1-\frac{2\epsilon
t}{\alpha})}{r^2}\bigg)=U_{eff}.\label{m1}
\end{align}

The effective potential extreme values are obtained by
$\frac{dU_{eff}}{dr}=0.$ The convolution point of effective
potential lies in the inner most circular orbit (ISCO).
\begin{equation}
r_0=\frac{L}{2M}\left(\left(\frac{t \epsilon}{\alpha}-1\right)\pm
\left(1-\frac{\epsilon t}{\alpha}\right)
\sqrt{(L^2+12M^2)-(L^2+9M^2)\frac{4 t \epsilon}{\alpha}} \right)
\end{equation}
The corresponding azimuthal angular momentum and the energy of the
particle at the ISCO are
\begin{align*}&L_{Z_0}^2=\frac{Mr(1+\frac{\epsilon t}{\alpha})}{1-\frac{3m}{r}},
\\&E^2_{0}=\frac{(r-2M)^2(1-\frac{\epsilon t}{\alpha})}{r(r-3m)}.\end{align*}

Consider the circular orbit $r = r_0$ of a particle, where $r_0$ is
local minima of effective potential. This orbit exist for $r_0 \in
(4M,\infty)$. The convolution point of effective potential for ISCO
is defined by $r_0=4M$ \cite{18}. Now suppose that the particle
collides with another particle which is in ISCO. There are three
possibilities after collision \textbf{(i)} bounded around BH,
\textbf{(ii)} captured by BH and \textbf{(iii)} escape to $\infty$.
The results depend upon the process of collision. Orbit of particle
is slightly change but remains bounded for small changes in energy
and momentum. Otherwise, it can be moved from initial position and
captured by BH or escape to infinity. After collision, the energy
and both angular momentum (total and azimuthal) change \cite{3}.
Before simplifying the situation, one can apply some conditions,
i.e., azimuthal angular momentum and initial radial velocity do not
change but energy can change by which we determine the motion of the
particle. Hence the effective potential becomes
\begin{align}{E}^2=\big(1-\frac{2M}{r})\bigg((1-\frac{\epsilon t}{\alpha})+
\frac{(L_z+rv_{\bot})^2(1-\frac{2\epsilon t}{\alpha})}{r^2\sin^2\theta}\bigg)=U_{eff}.\label{m2}\end{align}
This energy is greater then the energy of the particle before
collision, because after collision the colliding particle gives some
of its energy to the orbiting particle. Simplifying the Eq.
(\ref{m2}), we can obtain the escape velocity as follows
\begin{align}v_{\bot}=\bigg(\frac{rE^2(1+\frac{2\epsilon t}{\alpha})-(r-2M)
(1+\frac{\epsilon t}{\alpha})}{r-2M}\bigg)^{\frac{1}{2}}-\frac{L_z}{r}.\label{v}\end{align}

\subsection{Behavior of Effective Potential of Neutral Particle}

We analyze the trajectories of effective potential and explain the
conditions on energy required for bound motion or escape to infinity
around Schwarzschild BH. Figure \textbf{1} represents the behavior
of effective potential of particle moving around the Schwarzschild
BH for different values of angular momentum. We can see from Figure
\textbf{1}, the maxima of effective potential for $L_z = 5, 10, 15$
 at $r \approx 0.5, 1, 1.5$ for time interval $[0, 0.5)$. Similarly, the minima of effective potential for $L_z
= 5, 10, 15$ at $r \approx 0.5, 1, 1.5$  for time interval
$(0.5,1]$. Hence, it is concluded that maxima and minima is shifting
forward for large values of angular momentum. Also the repulsion and
attraction of effective force depends upon time. For time interval
$[0, 0.5)$, there is strong attraction near the BH and it vanishes
when radial coordinate approaches to infinity. Moreover, for time
interval $(0.5,1]$ there is strong repulsion near the BH and it
vanishes for $r$ approaches to infinity. For $t=0.5$, we find the
equilibrium i.e. there is no attraction and repulsion in effective
potential. Hence, effective potential is shifting from attraction to
repulsion when time increases and shifting point of time is $t =
0.5$. Hence, it is concluded that effective potential is attractive
and repulsive with respect to time, maxima and minima of effective
potential shifting forward in radial coordinate for large values of
angular momentum.
\begin{figure}
\centering
\includegraphics[width=7cm]{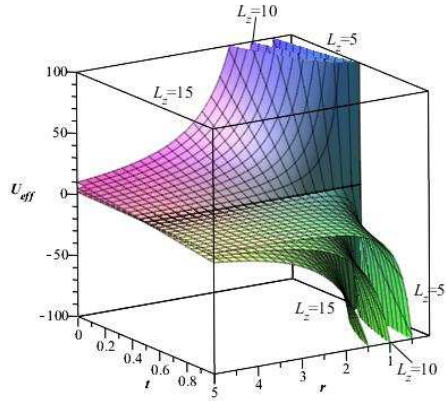}\\
\caption{The plot of effective potential ($U_\text{eff}$) versus $r$
and $t$ for $\alpha = \epsilon = 1$, $M = 10^{-16}$.}
\end{figure}
\begin{figure}[!htb]
  \centering
  \includegraphics[width=10cm]{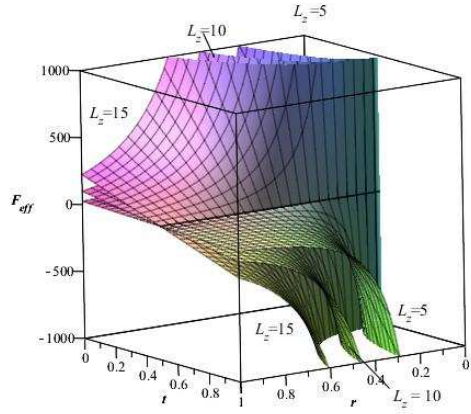}
  \caption{Plot of effective force versus $r$ and $t$ for $\alpha = \epsilon = 1$, $M = 10^{-16}$.}
\end{figure}
\begin{figure}[H]
  \centering
  \includegraphics[width=7cm]{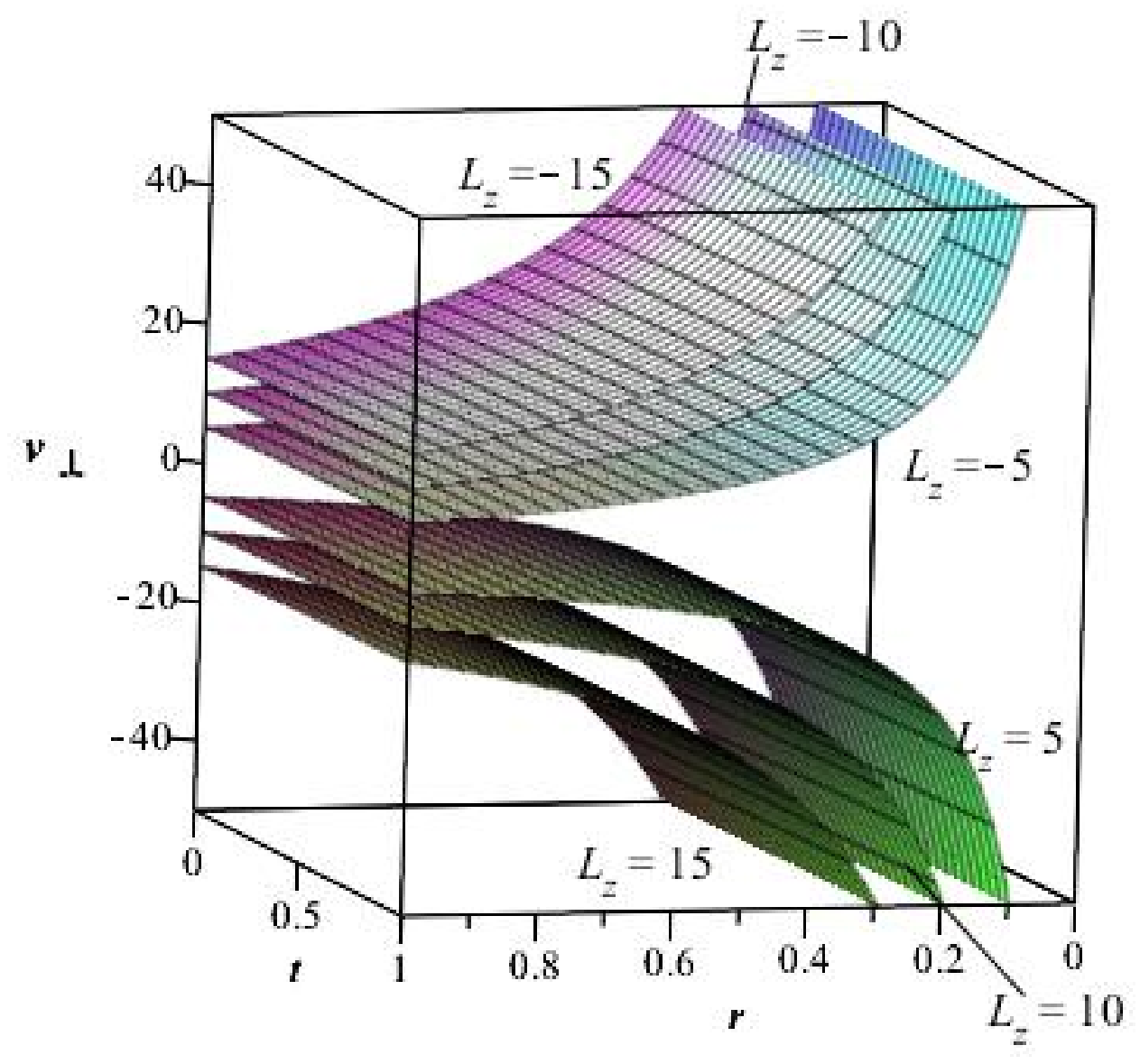}
  \caption{Plot of escape velocity versus $r$ and $t$ for $E = \alpha = \epsilon = 1$, $M = 10^{-16}$.}
\end{figure}

\subsection{Behavior of Effective Force of Neutral Particle}
The effective force is
\begin{align}F_{eff}=-\frac{1}{2}\frac{dU_{eff}}{dt}=
-\frac{M}{r^2}\bigg(1-\frac{\epsilon
t}{\alpha}+\frac{L_z^2\big(1-2\frac{\epsilon t}
{\alpha}\big)}{r^2}\bigg)+\big(1-\frac{2M}{r}\big)\frac{L_z^2\big(1-2\frac{\epsilon
t} {\alpha}\big)}{r^3}.\label{f}\end{align}
 We are studying the
motion of neutral particle in surrounding of Schwarzschild BH where
attractive and repulsive gravitational forces produce by scalar
vector tensor field which prevents a particle to fall into
singularity \cite{20}. The comparison of effective force on particle
around Schwarzschild BH as function of radial coordinate for
different values of angular momentum is shown in Figure \textbf{2}.
We can see from figure, at time interval $[0, 0.5)$, the attraction
of particle to reach singularity is more for $L_z=15$ as compared to
$L_z = 5, 10$. Similarly, at time interval $(0.5, 1]$, the repulsion
of particle to reach the singularity is more for $L_z=15$ as
compared to $L_z=5, 10$. Also, at $t=0.5$ there is no effective
force. Since $t=0.5$ acts as a shifting point, i.e., the effective
force is shifting from attraction to repulsion at $t=0.5$. We can
conclude that the attraction and repulsion of particle to reach and
escape from singularity is more for large values of angular momentum
but it mainly depend upon time.

\subsection{Trajectories of Escape Velocity of Neutral Particle}

Figure \textbf{3} represents the trajectories of escape velocity for
different values of $L_z$ as a function of $r$. It is evident that
if angular momentum is positive then escape velocity is repulsive
while it is attractive for negative values of angular momentum.
Also, the repulsion of escape velocity for large values of $L_z$ is
strong as compared to small values of $L_z$. On the other hand, the
attraction of escape velocity for $L_z = -15$ is strong as compared
to $L_z=-5, -10$. We can conclude that escape velocity of particle
increases as angular momentum increases but it becomes almost
constant away from the BH. Also it is interesting that the
trajectories of escape velocity do not vary with respect to time.

\section{Dynamics of the Charge Particle}

Now we consider the dynamics of charged particle around time
conformal Schwarzschild BH defined by (\ref{1}). We assume that a
particle has electric charge and its motion is affected by magnetic
field in BH exterior. Also, we assume that there exits magnetic
field of strength $(\mathbb{B})$ in the neighborhood of BH which is
homogenous, static and axisymmetric at spatial infinity. Next, we
follow the procedure of \cite{21} to construct magnetic field. Using
metric (\ref{1}), the general killing vector is
\cite{22}\begin{align}\Box \xi^{\mu}=0,\end{align} where $\xi^{\mu}$
is killing vector. Using above equation in Maxwell equation for 4-
potential $A^{\mu}$ in lorentz gauge $A^{\mu}_{;\mu}=0$, we have
\cite{23}
\begin{align}A^{\mu}
=\frac{\mathbb{B}}{2}\xi^{\mu}_{(\phi)},\label{u1}\end{align} The
killing vectors correspond to 4-potential which is invariant under
symmetries as follows \begin{align}L_{\xi}A_{\mu} = A_{\mu,\nu}
\xi^{\nu}+A_{\nu}\xi^{\nu}_{,\mu}=0.\end{align} Using magnetic field
vector \cite{16}
\begin{align}\mathbb{B}^{\mu} =
-\frac{1}{2}e^{\mu\nu\lambda\sigma}
F_{\lambda\sigma}u_{\nu},\label{u2}
\end{align}
where
$e^{\mu\nu\lambda\sigma}=\frac{\epsilon^{\mu\nu\lambda\sigma}}{\sqrt{-g}}$,
$\epsilon_{0123}=1$, $g=det(g_{\mu\nu})$,
$\epsilon^{\mu\nu\lambda\sigma}$ is Levi Civita symbol. The Maxwell
tensor is
\begin{align}F_{\mu\nu} =
A_{\nu,\mu}-A_{\mu,\nu}.
\end{align} In metric (\ref{1}), for local
observer at rest, we have
\begin{align} u^{\mu}_{0} =
\big(1-\frac{2M}{r}\big)^{-\frac{1}{2}} \xi^{\mu}_{(t)}, u^{\mu}_{3}
= (r sin\theta)^{-1}\xi^{\mu}_{(\phi)}.
\end{align}
The remaining two
components are zero at $\dot{r}=0$ (turning point). From Eqs.
(\ref{u1}) and (\ref{u2}), we get
\begin{align} \mathbb{B}^{\mu}
=\mathbb{B} \big(1-\frac{2M}{r}\big)^{-\frac{1}{2}}\left(cos\theta
\delta^{\mu}_{r}-\frac{sin\theta \delta^{\mu}_{\theta}}{r}\right),
\end{align} here the magnetic field
is directed along z-axis (vertical direction). Since the field is
directed upward, so we have $\mathbb{B}>0$.

The Lagrangian of particle moving in curved space time is given by
\cite{24} \begin{align}L = \frac{1}{2}g_{\mu
\nu}u^{\mu}u^{\nu}+\frac{qA}{m}u^{\mu},\end{align} where $m$ and $q$
are the mass and electric charge of the particle respectively. The
generalized 4-momentum of particle is
\begin{align}P_{\mu}=m u_{\mu}+q A_{\mu}.\end{align} The new
constants of motion are defined as
\begin{align}\dot{t}=\frac{E(1-\frac{\epsilon
t}{\alpha})}{\big(1-\frac{2M}{r}\big)},\quad
\dot{\phi}=\frac{L_z(1-\frac{\epsilon
t}{\alpha})}{r^2\sin^2\theta}-B,\label{u6}\end{align} where
$B=\frac{q \mathbb{B}}{2m}$. Using the constraints in the Lagrangian
(\ref{5}) the dynamical equations for $\theta$ and $r$ become
\begin{align}&\ddot{\theta}=\frac{\cos\theta(L_z(1-\frac{\epsilon t}
{\alpha})-B)^2}{r^4\sin^3\theta}-2\frac{\epsilon
\dot{\theta}Er}{\alpha(r-2M)}
-4\frac{\dot{r}\dot{\theta}}{r},\\&\ddot{r}=\frac{3M\dot{r}^2}{r(r-2M)}-
\frac{2\epsilon
Er\dot{r}}{\alpha(r-2M)}-\frac{ME^2(1-\frac{2\epsilon t}
{\alpha})}{r(r-2M)}+(r-2M)\bigg(\dot{\theta}^2+\sin^2\theta\big(\frac{L_z(1-
\frac{\epsilon
t}{\alpha})}{r^2\sin^2\theta}-B\big)^2\bigg).\label{u3}\end{align}
Using the normalization condition, we obtain
\begin{align}E^2=\dot{r}^2+r^2\big(1-\frac{2M}{r}\big)
\dot{\theta}^2+\big(1-\frac{2M}{r}\big)\bigg((1-\frac{\epsilon
t}{\alpha})+r^2\sin^2\theta\big( \frac{L_z(1-\frac{\epsilon
t}{\alpha})}{r^2\sin^2\theta}-B\big)^2\bigg).\end{align} The
effective potential takes the form
\begin{align}U_{eff}=\big(1-\frac{2M}{r}\big)\bigg((1-\frac{\epsilon t}
{\alpha})+r^2\sin^2\theta\big(\frac{L_z(1-\frac{\epsilon
t}{\alpha})}{r^2\sin^2\theta}-B\big)^2\bigg).\end{align} The energy
of the particle moving around the BH in orbit $r$ at the equatorial
plane is
\begin{align}U_{eff}=\big(1-\frac{2M}{r}\big)\bigg((1-\frac{\epsilon
t}{\alpha})+r^2\big(\frac{L_z(1-\frac{\epsilon
t}{\alpha})}{r^2}-B\big)^2\bigg).\label{U}\end{align} This is a
constraint equation i.e. it is always valid if it is satisfied at
initial time. Let us discuss the symmetric properties of
Eq.(\ref{u3}) which are invariant under the transformation as
follows
\begin{align}\phi\rightarrow - \phi, L_z \rightarrow -L_z, B
\rightarrow -B.\label{u4}\end{align} Therefore, without the loss of
generality, we consider $B>0$ and for $B<0$ we should apply
transformation (\ref{u4}) because negative and positive both charges
are inter related by the above transformation. However, if one
choose positive electric charge $(B>0)$ then both cases of $L_z$
(positive and negative) must be studied. They are physically
different: the change of sign of $L_z$ corresponds to the change in
direction of Lorentz force acting on the particle \cite{16}.

The system (\ref{u6}) - (\ref{u3}) is also invariant with respect to
reflection $\theta \rightarrow \pi - \theta.$ This transformation
preserves the initial position of the particle and changes
$v_{\bot}\rightarrow -v_{\bot}.$ Therefore it is sufficient to
consider the positive value of escape velocity $(v_{\bot})$
\cite{16}. By differentiating Eq.(\ref{U}) with respect to $r$, we
obtain
\begin{align}\frac{dU_{eff}}{dr}=\frac{2M}{r^2}\bigg(1-\frac{\epsilon
t}{\alpha}+r^2\big(\frac{L_z(1-\frac{\epsilon
t}{\alpha})}{r^2}-B\big)^2\bigg)-2rB\big(1-\frac{2M}{r}\big)\bigg(\frac{L_z(1-\frac{\epsilon
t}{\alpha})}{r^2}-B\bigg).\label{21}\end{align}

The effective potential after the collision when the body in the
magnetic field ($\theta=\frac{\pi}{2}$ and $\dot{r}=0$) is
\begin{align}E^2=U_{eff}=f(r)\bigg((1-\frac{\epsilon
t}{\alpha})+r^2\big(\frac{(L_z+v_{\bot}r)(1-\frac{\epsilon
t}{\alpha})}{r^2}-B\big)^2\bigg).\label{23}\end{align}
The escape
velocity of the particle takes the form
\begin{align}v_{\bot}=\bigg[\bigg(\frac{rE^2-(r-2M)(1-\frac{\epsilon
t}{\alpha})}{r-2M}\bigg)^{\frac{1}{2}}+rB\bigg](1+\frac{\epsilon
t}{\alpha})-\frac{L_z}{r}.\label{ve}\end{align} Differentiating
Eq.(\ref{21}) again with respect to $r$ we
have\begin{align}\frac{d^2U_{eff}}{dr^2}=\frac{2M}{r^2}\big(1-\frac{\epsilon
t}{\alpha}\big)+(4M-r)\bigg(\frac{L_z(1-\frac{\epsilon
t}{\alpha})}{r^2}-B\bigg)^2.\label{22}\end{align}

\subsection{Behavior of Effective Potential of Charged Particle}

\begin{figure}[H]
  \centering
  \includegraphics[width=7cm]{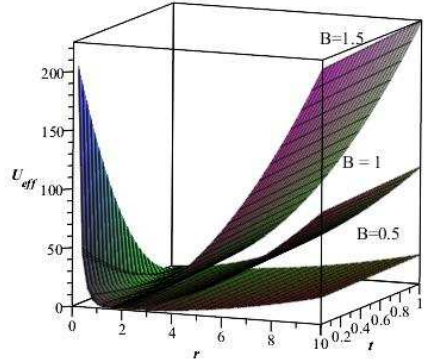}\\
  \caption{Graph of effective potential versus $r$ and $t$ given in Eq.(\ref{U}) for $\alpha = \epsilon = 1$, $M = 10^{-16}$ and $L_z = 3$.}
\end{figure}
Figure \textbf{4} represents the behavior of effective potential as
a function of radial coordinate for different values of magnetic
field \emph{B}. The minima of effective potential at $B=0.5$ is
approximately at $r=2$ initially but as time increase it shifting
near the BH. Hence the presence of magnetic field increases the
possibility of particle to move in stable orbit. Similarly,
initially the minima of effective potential at $B=1, 1.5$ is
approximately at $r=1.5, 1.25$ respectively, as time increases it
approaches near BH. One can noticed that in the presence of low
magnetic field, the minima of effective potential is shifted away
from the horizon and width of ISCO is also decreased as compared to
high magnetic field. These results is an agreement with
\cite{21,25}. Therefore we can say that increase in magnetic field
act as increase instable orbits of particle. Another aspect from
figure is that at initial time, we have possibility of stable orbit
but as time increases the possibility of ISCO getting low and at
$t=1$, the stable orbit is lost. Hence it is possible that the
particle is captured by the BH or it escape to infinity.

\subsection{Behavior of Effective Force of Charged Particle}
The effective force on the particle can be defined as
as\begin{align}F_{eff}=-\frac{1}{2}\frac{dU_{eff}}{dr}=-\frac{M}{r^2}\bigg(1-\frac{\epsilon
t}{\alpha}+r^2\big(\frac{L_z(1-\frac{\epsilon
t}{\alpha})}{r^2}-B\big)^2\bigg)+rB\big(1-\frac{2M}{r}\big)\bigg(\frac{L_z(1-\frac{\epsilon
t}{\alpha})}{r^2}-B\bigg).\label{f1}\end{align}
\begin{figure}[H]
  \centering
  \includegraphics[width=8cm]{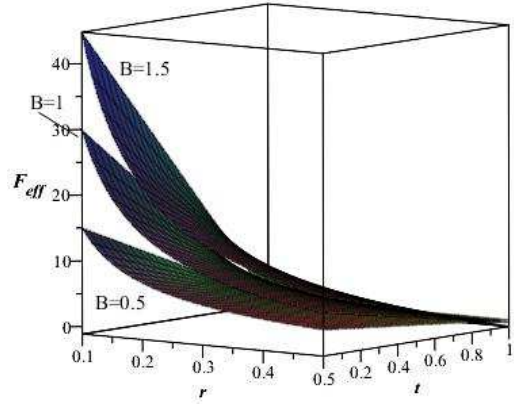}\\
  \caption{Graph of effective force versus $r$ and $t$ given in Eq. (\ref{f1}) for $\alpha = \epsilon = 1$, $M = 10^{-16}$ and $L_z = 3$.}
\end{figure}
We have plotted the effective force for different values of B as
function of $r$ . We see from Figure \textbf{5} that effective force
is more attractive for large values of magnetic field as compared to
small values. We can conclude that effective force of particle
increases as strength of magnetic field increases near BH initially
but with the passage of time, the particle moves away from BH it
becomes almost constant.

\subsection{Trajectories of Escape Velocity of Charged Particle}
From Eq.(\ref{u6}), the angular variable we have
\begin{align}\dot{\phi}=\frac{L_z(1-\frac{\epsilon
t}{\alpha})}{r^2\sin^2\theta}-B.\end{align} If the the Lorentz force
on particle is attractive then left hand side of above equation is
negative \cite{26} and vice verse. The motion of charged particles
is in clockwise direction. Our main focus on magnetic field acting
on particle, the large value of magnetic field deforms the orbital
motion of particle as compared to small values. Hence, we can
conclude that the possibility to escape the particle from ISCO is
greater for large values of magnetic field. We also explain the
behavior of escape velocity for different values of magnetic field
in Figure \textbf{6} graphically. The escape velocity of particle
increases for large values of magnetic field as compared to small
values. We can conclude that presence of magnetic field provide more
energy to particle to escape from vicinity of BH. Figure \textbf{7}
represents escape velocity against $r$ for different values of
angular momentum. The possibility of particle to escape is low for
large values of angular momentum. Also as time increases the
possibility of particle to escape is decreasing drastically.
\begin{figure}[H]
  \centering
  \includegraphics[width=8cm]{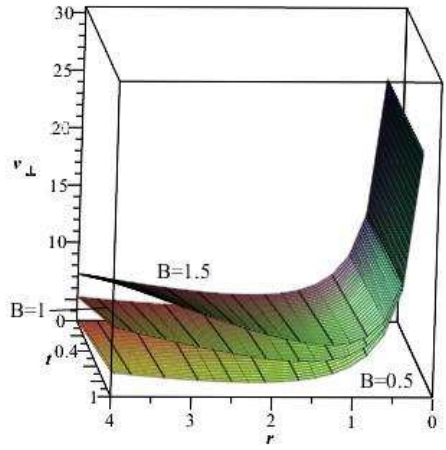}\\
  \caption{Plot of escape velocity versus $r$ and $t$ given in Eq.(\ref{ve}) for $E = \alpha = \epsilon = 1$, $M = 10^{-16}$ and $L_z = 5$.}
\end{figure}

\begin{figure}[H]
  \centering
  \includegraphics[width=8cm]{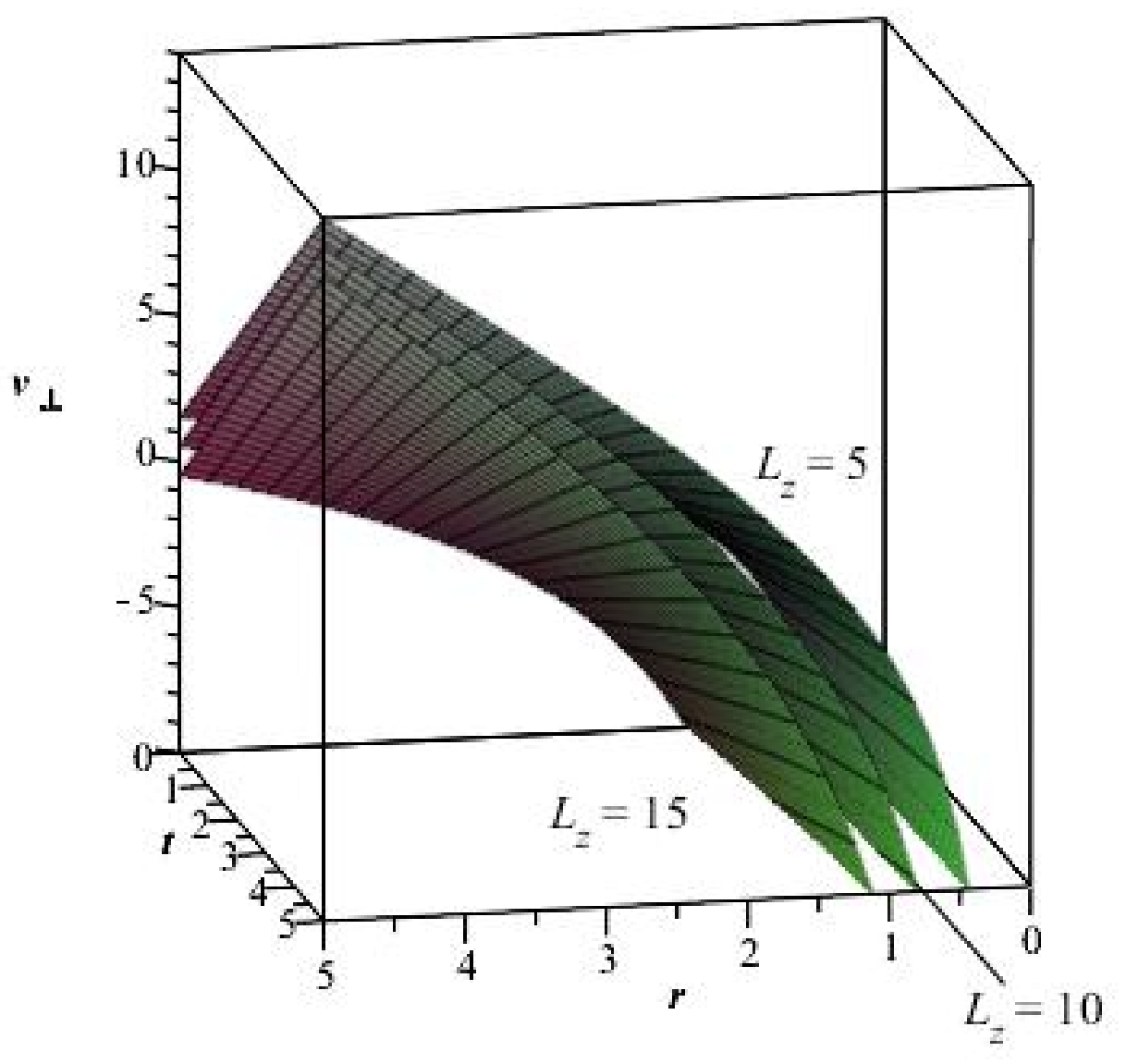}\\
  \caption{Plot of escape velocity versus $r$ and $t$ given in Eq.(\ref{ve}) for $E = \alpha = \epsilon = 1$, $M = 10^{-16}$ and $B=0.5$.}
\end{figure}
\section{Stability Orbit} The Lyapunov function is defined as
\cite{27}
\begin{align}\lambda=\bigg(\frac{-U_{eff}^{''}(r_0)}{2\dot{t}^2(r_0)}\bigg)^{\frac{1}{2}}.\end{align}Using Eq.(\ref{22}) we can get the
Lyapunov function
as\begin{align*}\lambda=&\frac{\big(1+\frac{\epsilon
t}{2\alpha}\big)(r-2M)}{r^{\frac{5}{2}}E}\bigg[2M\bigg(1+2L_z\big(\frac{(1-\frac{\epsilon
t}{\alpha})L_z}{r^2}-B\big)\bigg)+B\bigg(rL_z-r^3B(1+\frac{\epsilon
t}{\alpha})\\-&2L_z(r-2M)\bigg)\bigg]^{\frac{1}{2}}.\end{align*}
\begin{figure}[H]
\centering
\includegraphics[width=6cm]{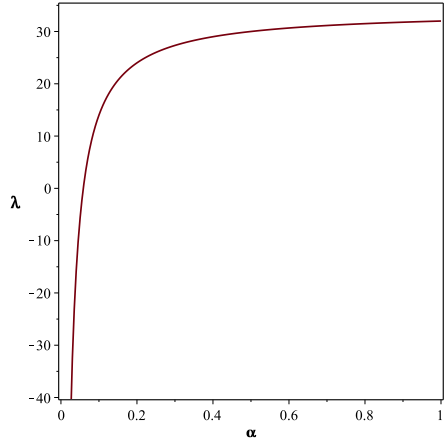}\\
\caption{Lyapunov exponent as a function for $\alpha$ for $B =
\epsilon = 1$, $M = 10^{-16}$, $L_z = ??$, $r = 1$and $t=1$.}
\end{figure}
Behavior of $\lambda$ as function of $\alpha$ is analyzed Figure
\textbf{8}. We can see from figure that instability of circular
orbit increases as $\alpha$ increases but it becomes constant

\section{Conclusion and Observations}

In the present work, we found that one of the Noether symmetry admit
approximated part. This symmetry is the translation in time which
means that the energy of the spacetime is re-scaling. We also found
the conservation laws corresponding to the exact Schwarzschild
spacetime and the time conformal Schwarzschild spacetime and compare
their numerical solutions. The geodesic deviation has also
presented. The perturbation in the spacetime has perturbed the
geodesic. The numerical solutions showed that how much they deviated
from each other.

Using Noetherine symmetries, we have calculated the equation of
motions. We have found three types of approximate Noether symmetries
which correspond to energy, scaling and Lorentz transformation in
the conformal plane symmetric spacetimes. These symmetries
approximate the corresponding quantities in the respective
spacetimes. We have not seen approximate Noether symmetries
corresponding to linear momentum, angular momentum and galilean
transformation in our calculations. This shows that these quantities
conserved for plane symmetric spacetimes. The spacetime section of
zero curvature does not admit approximate Noether symmetry
\cite{ali1} which shows that the approximate symmetries disappeared
whenever we have the section of zero curvature in the spacetimes.
However, the approximate Noether symmetry does not exist in flat
spacetimes (Minkowski spacetime).

In addition, we have investigated the motion of neutral and charged
particles in the absence and presence of magnetic field around the
time conformal Schwarzschild BH. The behavior of effective
potential, effective force and escape velocity of neutral and
charged particles with respect to time are discussed. In case of
neutral particle, for large values of angular momentum, we have
found attraction and repulsion with respect to time as shown in
Figures \textbf{1-3}. This effect decreases far away from BH with
respect to time. The more aggressive attraction and repulsion of
particle to reach and escape from BH has been observed for large
values of angular momentum due to time dependence. The escape
velocity increases as angular momentum increasing but it does not
vary with respect to time.

In case of charged particle, the presence of high magnetic field
shifted the minima of effective potential towards the horizon and
width of stable region is increased as compare to low magnetic
field. Escape velocity for different values of angular momentum and
magnetic field are shown in Figure \textbf{6-7}. It is found that
the presence of magnetic field provided more energy to particle to
escape. The possibility of particle to escape is low for large
angular momentum and as time increases it decreases continuously.

\end{document}